\documentclass[graybox]{svmult}

\usepackage{mathptmx}       % selects Times Roman as basic font
\usepackage{helvet}         % selects Helvetica as sans-serif font
\usepackage{courier}        % selects Courier as typewriter font
\usepackage{type1cm}        % activate if the above 3 fonts are
\usepackage{makeidx}         % allows index generation
\usepackage{graphicx}        % standard LaTeX graphics tool
\usepackage{multicol}        % used for the two-column index
\usepackage[bottom]{footmisc}% places footnotes at page bottom
\usepackage{natbib} 
\newcommand{\teff}{\ensuremath{T_{\mathrm{eff}}}}  
\newcommand{\logg}{\ensuremath{\log g}}
\newcommand{\feh}{\ensuremath{[\rm{Fe}/\rm{H}]}}

\makeindex     % used for the subject index
\begin{document}

\title*{Spectroscopic analysis of cool giants and supergiants}
%
% Use \titlerunning{Short Title} for an abbreviated version of
% your contribution title if the original one is too long
%
\author{Maria Bergemann, Rolf-Peter Kudritzki, and Ben Davies}
%
% Use \authorrunning{Short Title} for an abbreviated version of
% your contribution title if the original one is too long
%
\institute{M. Bergemann \at Institute of Astronomy, University of Cambridge, CB3
0HA, Madingley Road, UK\\\email{mbergema@ast.cam.ac.uk} \and Rolf-Peter
Kudritzki \at Institute for Astronomy, University of Hawaii, 2680 Woodlawn
Drive, Honolulu, HI 96822, US; Ludwig Maximilian University of Munich, Scheiner
str. 1, 81679, Germany\\ \email{kud@ifa.hawaii.edu} \and Ben Davies \at
Astrophysics Research institute, Liverpool John Moores University, IC2,
Liverpool Science Park, 146 Brownlow Hill, Liverpool L3 5RF, UK\\
\email{B.Davies@ljmu.ac.uk}}
%
% Use the package "url.sty" to avoid
% problems with special characters
% used in your e-mail or web address
%
\maketitle

\abstract*{}
\abstract{
Cool red giants and supergiants are among the most complex and fascinating stars
in the Universe. They are bright and large, and thus can be observed to enormous
distances allowing us to study the properties of their host galaxies, such as
dynamics and chemical abundances. This review lecture addresses various
problems related to observations and modelling spectra of red giants and
supergiants. }

\section{Introduction}
\label{sec:1}

\begin{figure}[!htb]
\sidecaption
\includegraphics[scale=.35]{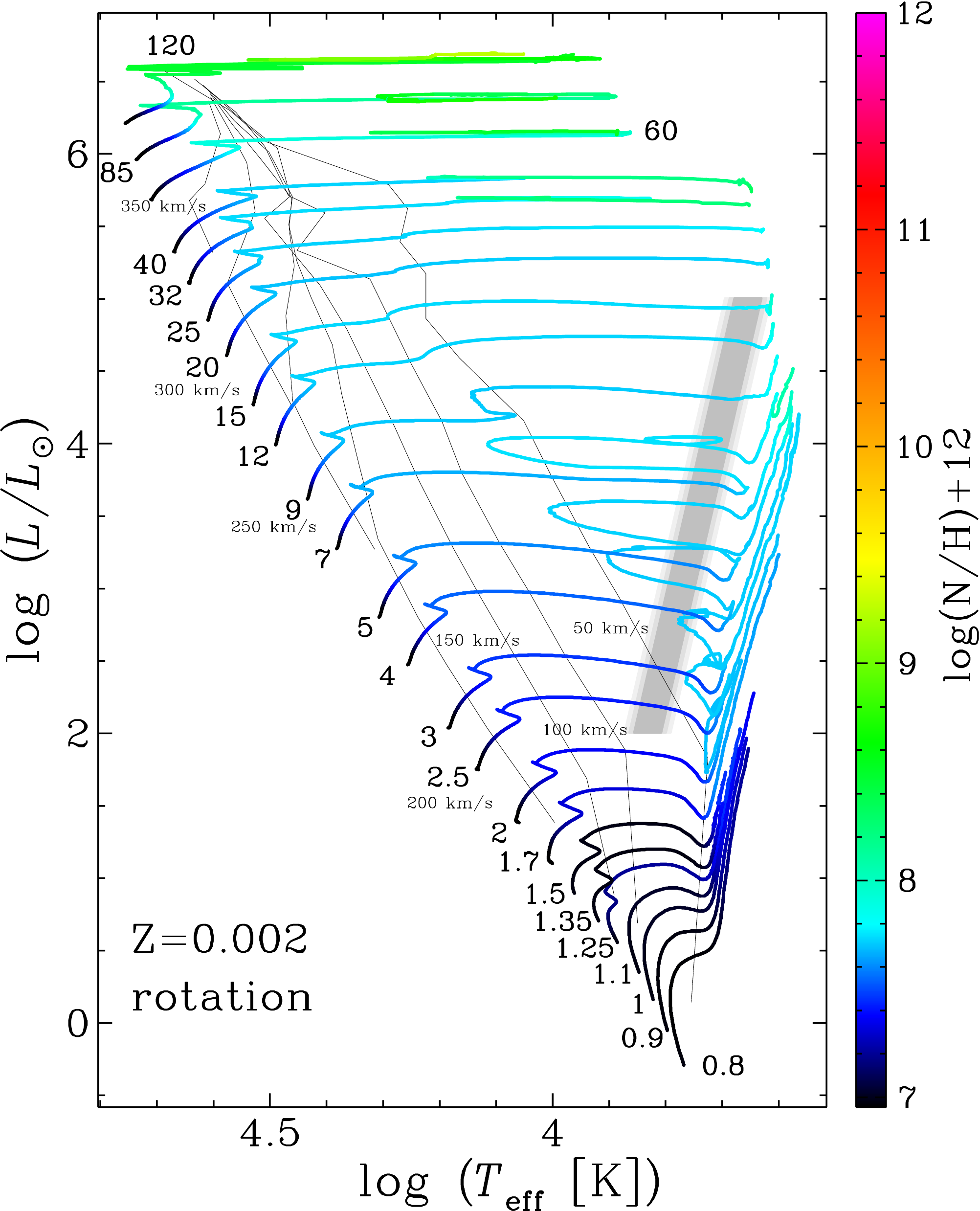}
\caption{Stellar tracks including the effect of rotation for stellar models with
initial mass of $0.8$ to $120$ $M_{\odot}$ \cite[][]{2013A&A...558A.103G}. Cool
giants and red
supergiants occupy the stripe extending vertically from $1 < \log (L/L_{\odot})
< 6$.}
\label{fig:1}   
\end{figure}

Cool evolved stars are perhaps the most enigmatic cosmic objects with
luminosities spanning several orders of magnitude. The stars are at the end
stages of stellar evolution (Fig. \ref{fig:1}) occupying the coolest vertical
strip on the Herzsprung-Russell diagram, the Hayashi limit for fully convective
stars.

Low- and intermediate-mass stars evolve to the red giant branch (RGB) and
asymptotic giant branch (AGB) after having spent most of their lifetime on the
main sequence. These stars are cool and luminous, with effective temperatures
$\teff$ between $2000$ and $5500$ K and  luminosities $L$ between $10$ and
$10^4$ $L_\odot$. The stars have a wide range of ages, from 1 to $> 10$ Gyr, and
thus trace chemical composition of interstellar matter in galaxies now and in
the past. 

High-mass stars are those with masses from $10$ to $60$ $M_\odot$; they evolve
and explode quickly. Red supergiants (RSG) are young, typically $< 50$ Myr, yet
extremely bright with $L$ from $10^4$ to $10^6 L_\odot$. 

Nucleosynthesis taking place in the interior of a star has as a consequence
that, as evolution proceeds, stellar atmosphere acquires abundance patterns that
strongly differ from the chemical composition of the natal cloud in which the
star was born. This is referred to as self-pollution: dredge-up episodes bring
material from the interior to the surface. Thus the stellar atmosphere becomes
enriched or depleted in different chemical elements, e.g., He, C and N, and
s-process elements \citep[e.g.][]{1992eatc.conf...92L, 1993ApJ...413..641V,
2005ARA&A..43..435H}. These abundance peculiarities are extreme on the AGB
phase. The newly synthesised elements are returned into the ISM, as stars lose
mass through winds. Giant stars are thus the primary producers of chemical
elements, and much can be learned about stellar evolution from the analysis of
abundance ratios as a function of the evolutionary stage of a star.

Giant stars serve as a primary gauge of cosmic abundances. They are the key
targets in studies of Galactic chemical evolution and stellar archeology.
Ongoing observational programmes search for very- and ultra-metal poor (UMP)
stars in the Galactic halo and in the bulge \citep{2005ARA&A..43..531B}; many of
such objects are old red giants. Some large-scale stellar surveys, such as
APOGEE (Apache Point Galactic Experiment from Sloan Digital Sky Survey) focus
entirely on RGB stars, for their intrinsic brightness allows to probe very large
range of distances in the Galaxy. Individual giant stars can be observed with
modern telescopes in the nearby galaxies of the Local Group
\citep[e.g.][]{2006AJ....131.2497G, 2010ApJS..191..352K}. Moreover, RGB and AGB
stars dominate the integrated light of spatially unresolved systems, such as
extra-Galactic globular clusters, dSph, and elliptical galaxies, thus allowing
us to determine chemical composition of stellar populations from their composite
spectra.

The atmospheres of giant stars are not fully in hydrostatic and thermodynamic
equilibrium. Theory can explain observations only if very complex physical
phenomena are included in the models: outflows, shocks, winds, pulsations,
indicating that we are dealing with stars far more complex than the Sun. Violent
convective motions penetrate their atmospheres and influence the physical state
of matter. The extremely low densities of giant photospheres are to be blamed
for the fact that matter and radiation are not in equilibrium.

All these phenomena will be briefly reviewed below \citep[for a status update
see also the review by][]{2010AN....331..433R}. First, we recap the most recent
results from imaging and spectroscopy of cool stars (Sec. \ref{sec:2}), which
convincingly show that the stars are more sophisticated compared to their
evolutionary predecessors, un-evolved main-sequence stars, and give a glimpse to
their complex physics. Sec. \ref{sec:3} describes the methods used for stellar
parameter analysis. In Sec. \ref{sec:4}, we review the key results of
state-of-the-art modelling of cool giant spectra\footnote{A more detailed
discussion of model atmospheres and synthetic spectra for giants are given in
the review lecture on 3D NLTE spectroscopy.}.
\section{Observations}
\label{sec:2}
With the recent developments in astronomical instrumentation, in particular
adaptive optics, it has become possible to obtain unique observations of giants
in their native spectral range, infra-red (IR)\footnote{Very cool stars, such
as RGB and RSG, emit the most light in the IR.}.
\begin{figure}[t]
\sidecaption
\hbox{
\includegraphics[scale=.3]{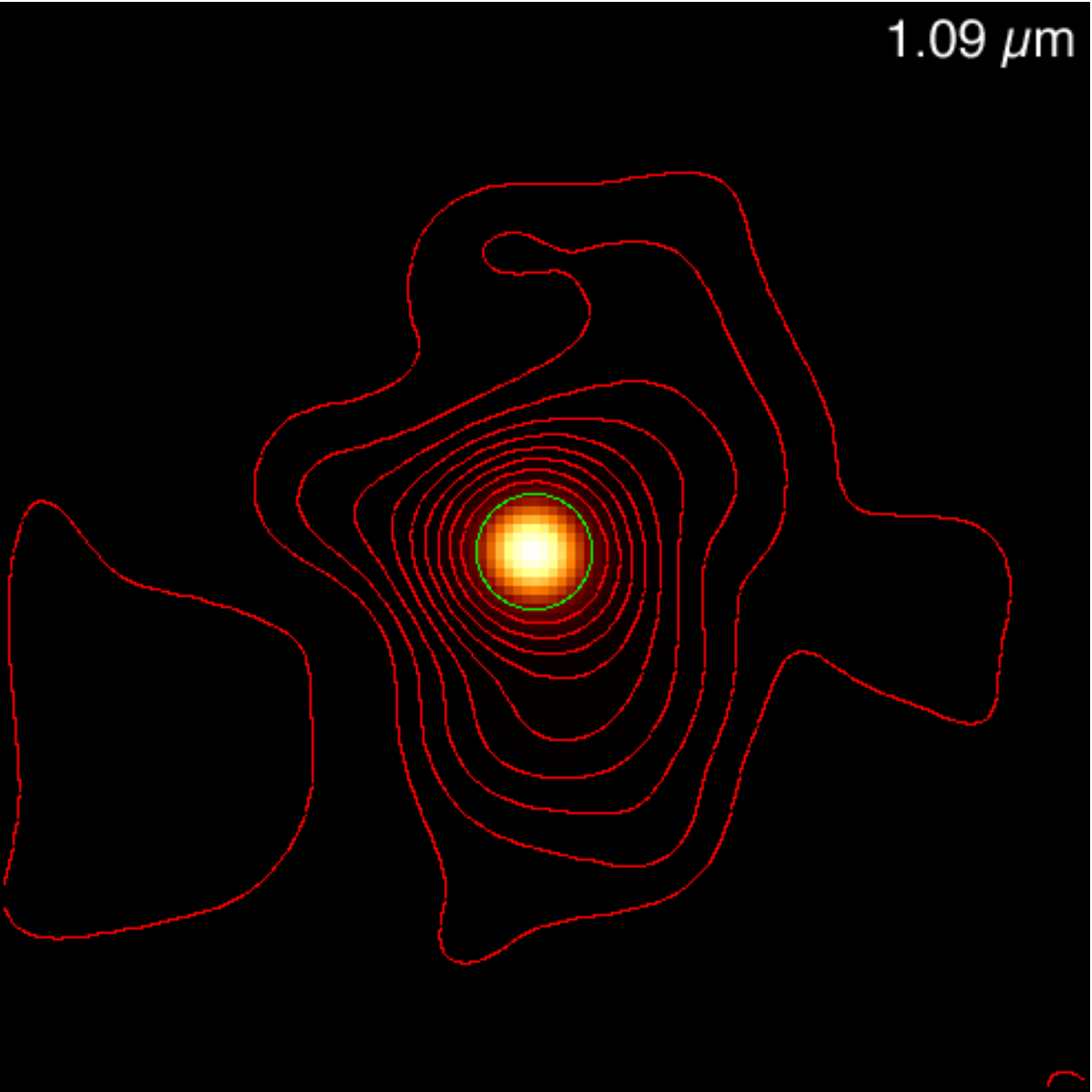}
\includegraphics[scale=.3]{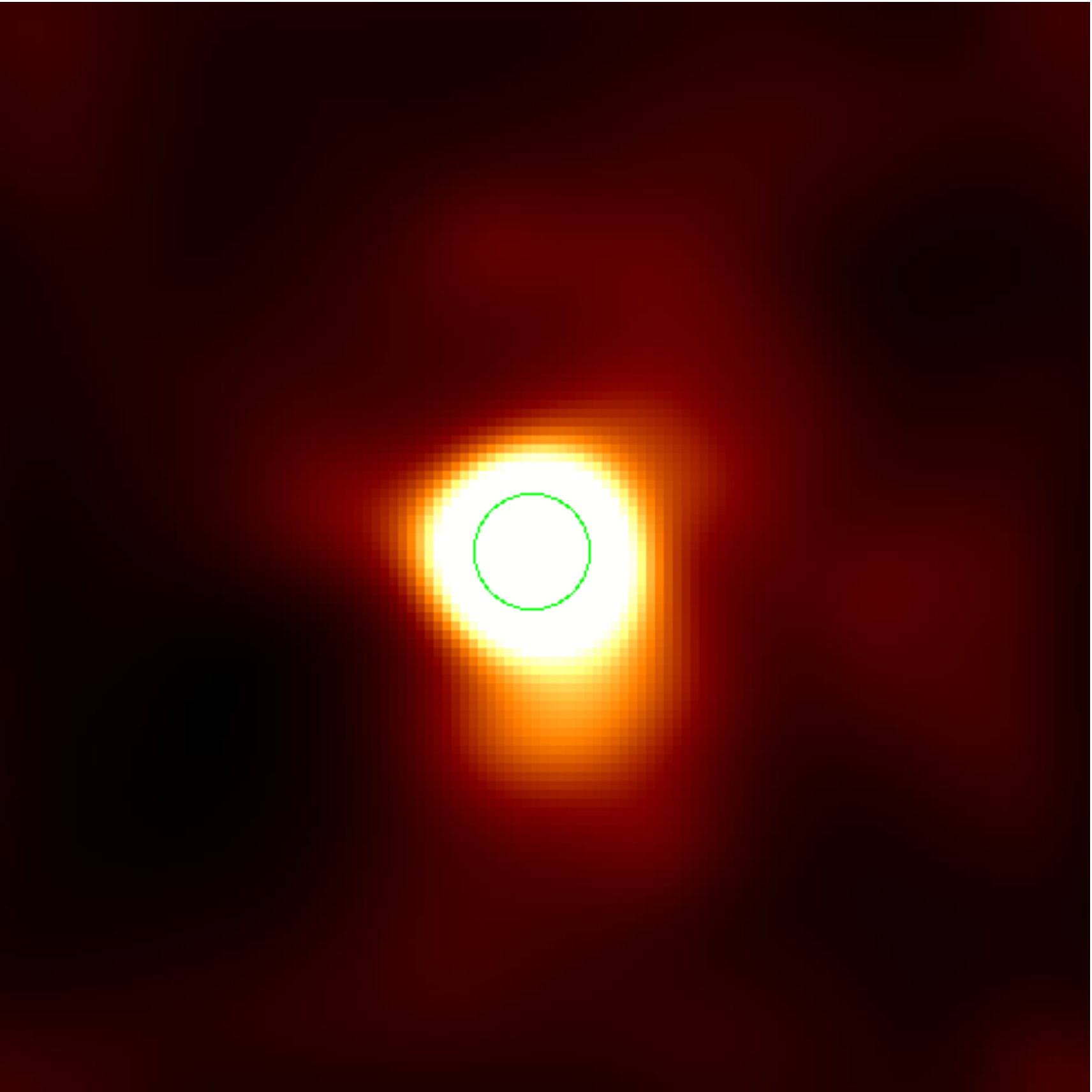}}
\caption{Interferometric observations of Betelgeuse taken with VLT/NACO at
$1.09\mu$m \citep[][]{2009A&A...504..115K}. The left image has a linear colour
scale; the contours indicate different flux levels in increasing powers of 2.
The right plot corresponds to a square-root colour scale and it also shows the
photospheric radius of the star at 43.7 mas (green circle).}
\label{fig:2}   
\end{figure}

Diffraction-limited imaging observations of giants and supergiants can be done
with ground-based interferometric telescopes, such as VLT/NACO
\citep{2009A&A...504..115K}.  Surface inhomogeneities on RSGs were detected in
the JHK bands. Betelgeuse, our closest red supergiant with a radius $R \sim 800 
R_{\odot}$,
appears to have a complex circumstellar envelope with an outflow, or plume,
extending to enormous distances of several stellar radii (Fig. \ref{fig:2}). A
combined analysis of different photometric bands suggests the presence of a very
cool layer on top of the Betelgeuse's photosphere, perhaps made of CN or water
molecules \citep{2009A&A...504..115K},  and dusty envelopes, which produce flux
excess in the far-IR
stellar spectral energy distributions \citep{2000ApJ...538..801T}.
Interferometric long-baseline observations in optical filters
\citep{1993ApJ...416L..25Q}, discovered another interesting effect: the star
imaged in different filters,
e.g. TiO $712$ nm vs continuum at 754 nm, has a different size! The difference
in diameter for the coolest M-type giants and supergiants is nearly 10 $\%$,
which suggests huge extension of their atmospheres.

\begin{figure}[b]
\sidecaption
\includegraphics[scale=.15]{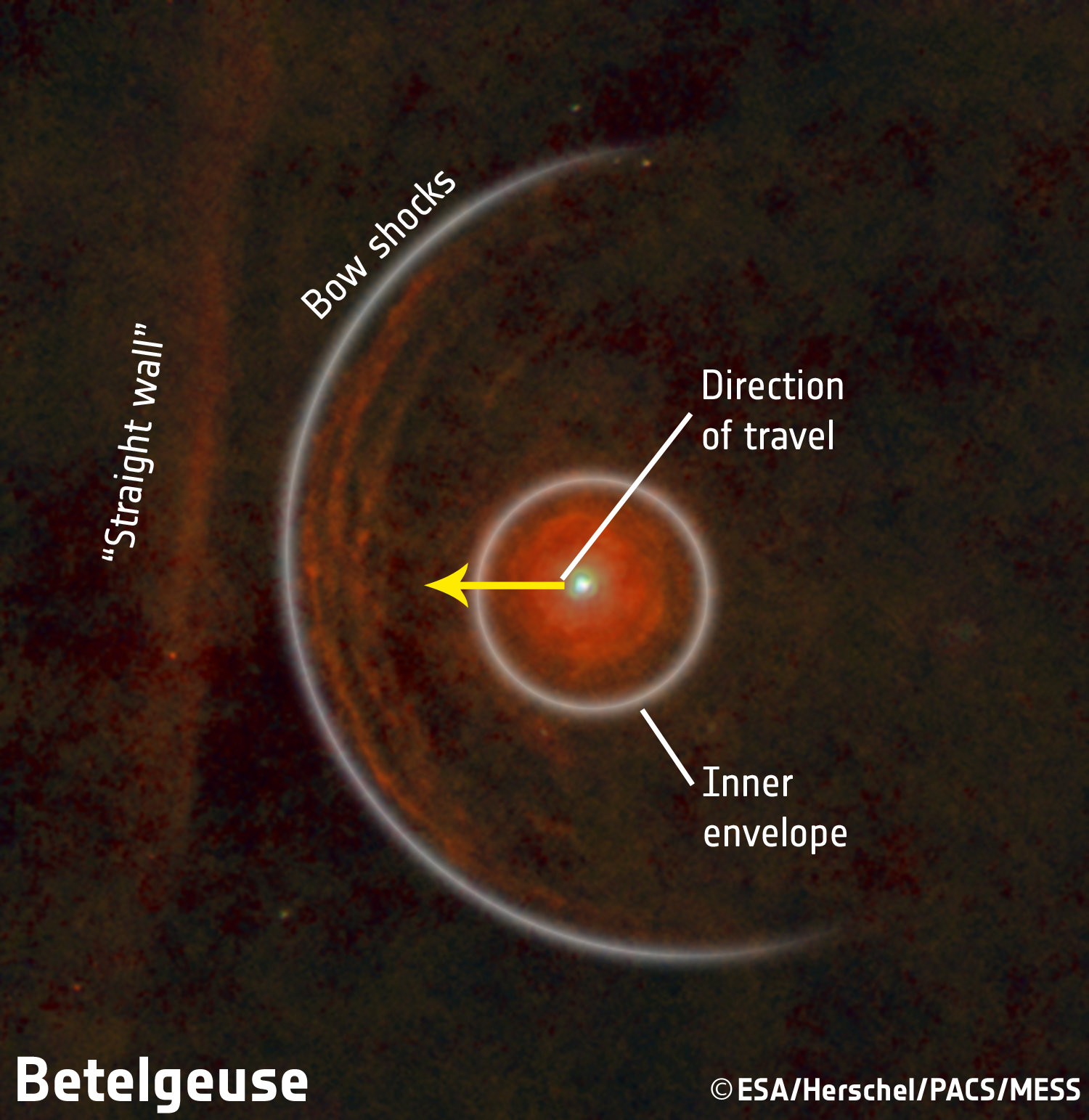}
\caption{Herschel image of Betelgeuse \cite[][]{2012A&A...548A.113D}. The image
shows the morphology of the stars's environment (see text).}
\label{fig:3}
\end{figure}

Other interesting results about atmospheric structure and dynamics of cool stars
were delivered by space missions, such as the Herschel Space Observatory
\citep[][HSO, e.g.]{2012A&A...545A..99T}. Fig. \ref{fig:3} shows the image of
Betelgeuse made by superposition of images taken at $70$ to $160$ $\mu$m
\citep{2012A&A...548A.113D}. The image reveals a dusty envelope and multiple
arc-like structures caused by the interaction of stellar wind and interstellar
matter, so-called bow shocks. The morphology of these structures can be
explained by sophisticated 3D time-dependent hydrodynamical simulations of the
stellar envelopes.

\begin{figure}[!htb]
\sidecaption
\includegraphics[scale=.45]{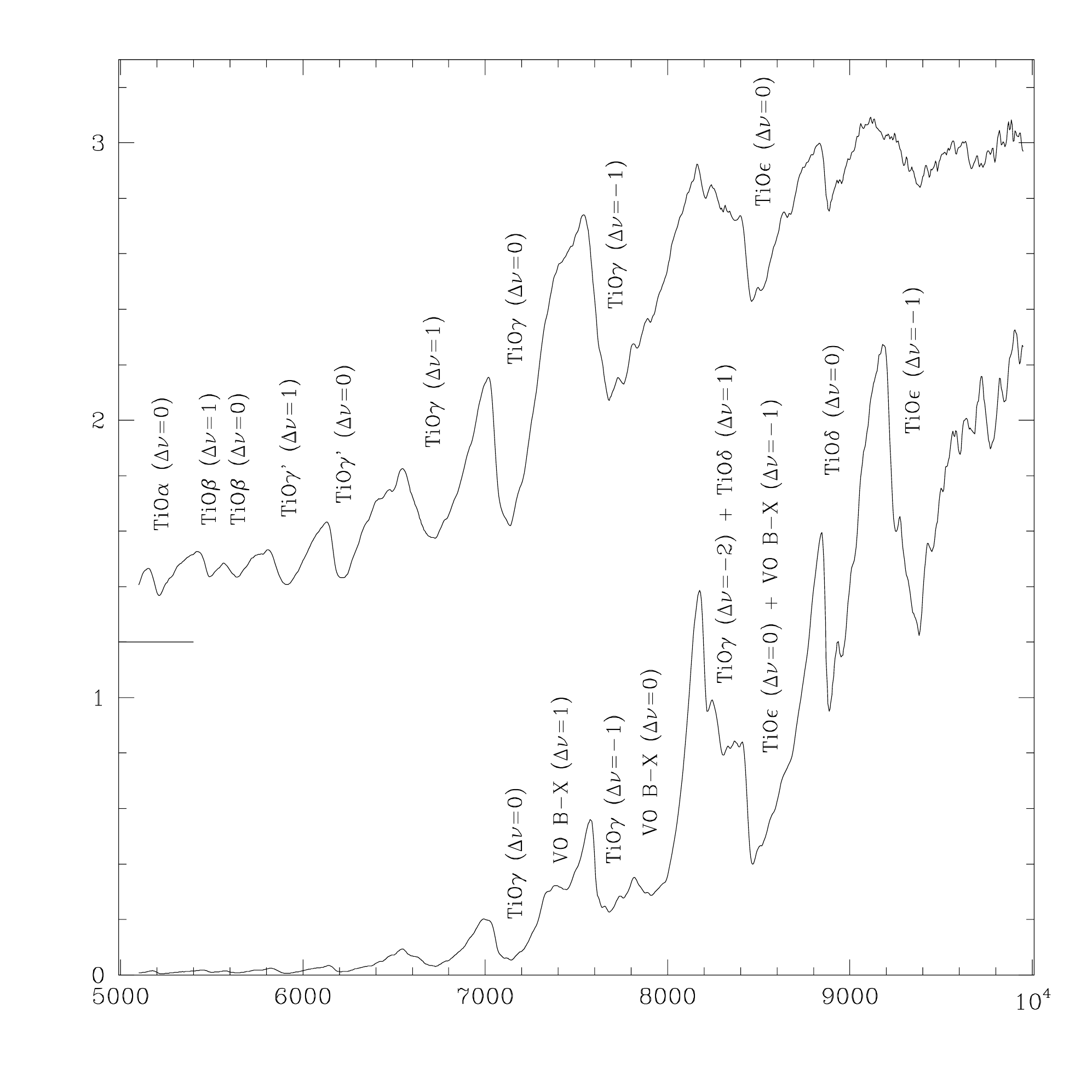}
\includegraphics[scale=.45]{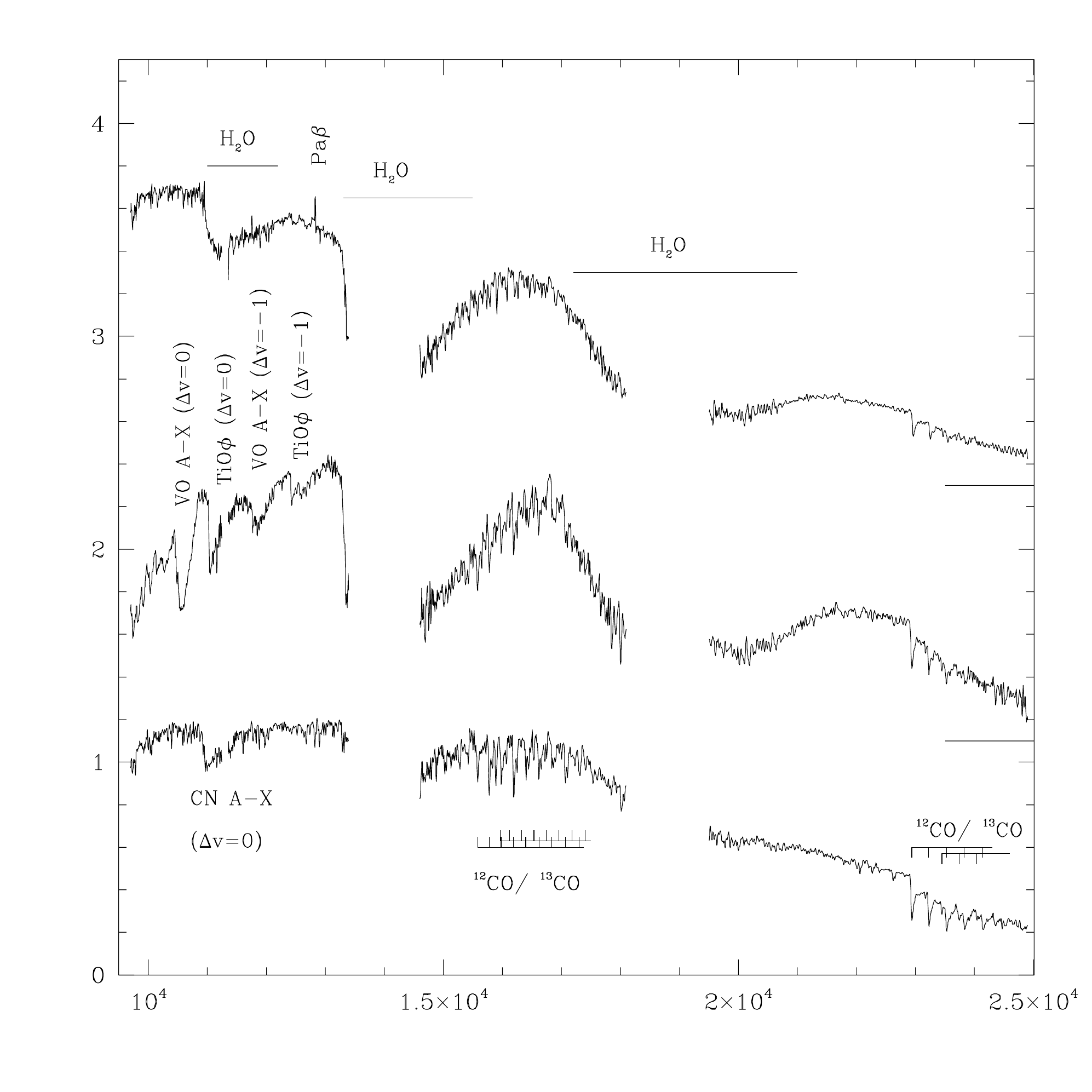}
\caption{Spectra of cool stars \cite[][]{2000A&AS..146..217L}. Top: a spectrum
of a warm Mira star in the optical window below $1 \mu$m; the TiO bands are
prominent. Bottom: the IR spectra of a warm Mira star, cool Mira star, and the
red supergiant (top to bottom).}
\label{fig:4}   
\end{figure}

Spectroscopic observations reveal complexities of a different type. Red giants
and supergiants are very cool stars, with effective temperatures from $\sim
3000$ to $5000$ K. In these conditions, various molecules very efficiently
absorb radiation. Thus, the spectra are severely distorted by molecular bands,
such as CN, NH, H$_2$O, and particularly TiO \citep{2000ApJ...540.1005A}, which
absorbs most of the radiation in the optical window (Fig. \ref{fig:4} top
panel). Although molecules are ubiquitous and are sensitive probes of
atmospheric physics, their modelling is very difficult (Sec \ref{sec:4}) and
complex 3D radiative hydrodynamics calculations are  needed
\citep{2011ApJ...736...69U}. 

Very high-resolution observations of cool giants
\citep[e.g.][]{1990A&A...228..218D, 2008AJ....135.2033G, 2010ApJ...725L.223R,
2012AJ....143..112G} show that the spectral lines are not symmetric. The
reversed-$C$ shapes of the line profiles indicate large convection cells at the
surface of stars.

The IR part of giant spectra is much less contaminated by molecular absorption
that significantly simplifies the problem of spectral analysis. Targeting the
near-IR region also has the benefit that the stellar flux is less extinguished
by interstellar absorption, allowing studies deep into the galaxy. For example,
cool giants and supergiants have been used to study the abundances in the
Galactic Centre  \citep{2000ApJ...530..307C, 2007ApJ...669.1011C, 
2009ApJ...696.2014D}, the Scutum Arm tangent and the end of the Galactic Bar
\citep{2009ApJ...694...46D}, and the Galactic Bulge
\citep{2007ApJ...665L.119R}. 

Thus in terms of science, spectroscopy of RGB, AGB, and RSG stars has a great
potential in the IR.
\section{Stellar parameters}\label{sec:3}
The basic parameters of cool stars can be estimated by different methods. Here,
we focus on the spectroscopic techniques that have been developed to determine
effective temperature $\teff$, surface gravity $\log g$, and metallicity $\feh$
of a star.

The perhaps most stable and simple method to estimate stellar parameters is the
method of excitation-ionization balance for a pair of chemical species, such as
Fe I and Fe II. The simplicity comes mainly from the application of the
curve-of-growth technique \citep{2008oasp.book.....G}.  First, one computes the
abundance from the measured line equivalent width. The $\teff$ is then set by
minimising the magnitude of the slope of the relationship between the abundance
of iron from Fe I lines and the excitation potential of each line. The surface
gravity is then estimated by minimising the difference between the abundance of
iron measured from Fe I and Fe II lines. Iterations are needed to have the above
criteria satisfied. The method offers several advantages: efficiency, fast
convergence, and easy implementation. 
The downside is the need for a statistically significant sample of clean
unblended lines; moreover, the method is very sensitive to the physics of the
models. Recent studies showed that standard 1D hydrostatic model atmospheres and
LTE line formation fail for giant stars, producing unphysically low estimates of
temperature, surface gravity, and metallicity
\citep[e.g.][]{2012MNRAS.427...27B, 2013MNRAS.429..126R}.

\begin{figure}[b]
\sidecaption
\includegraphics[scale=.2]{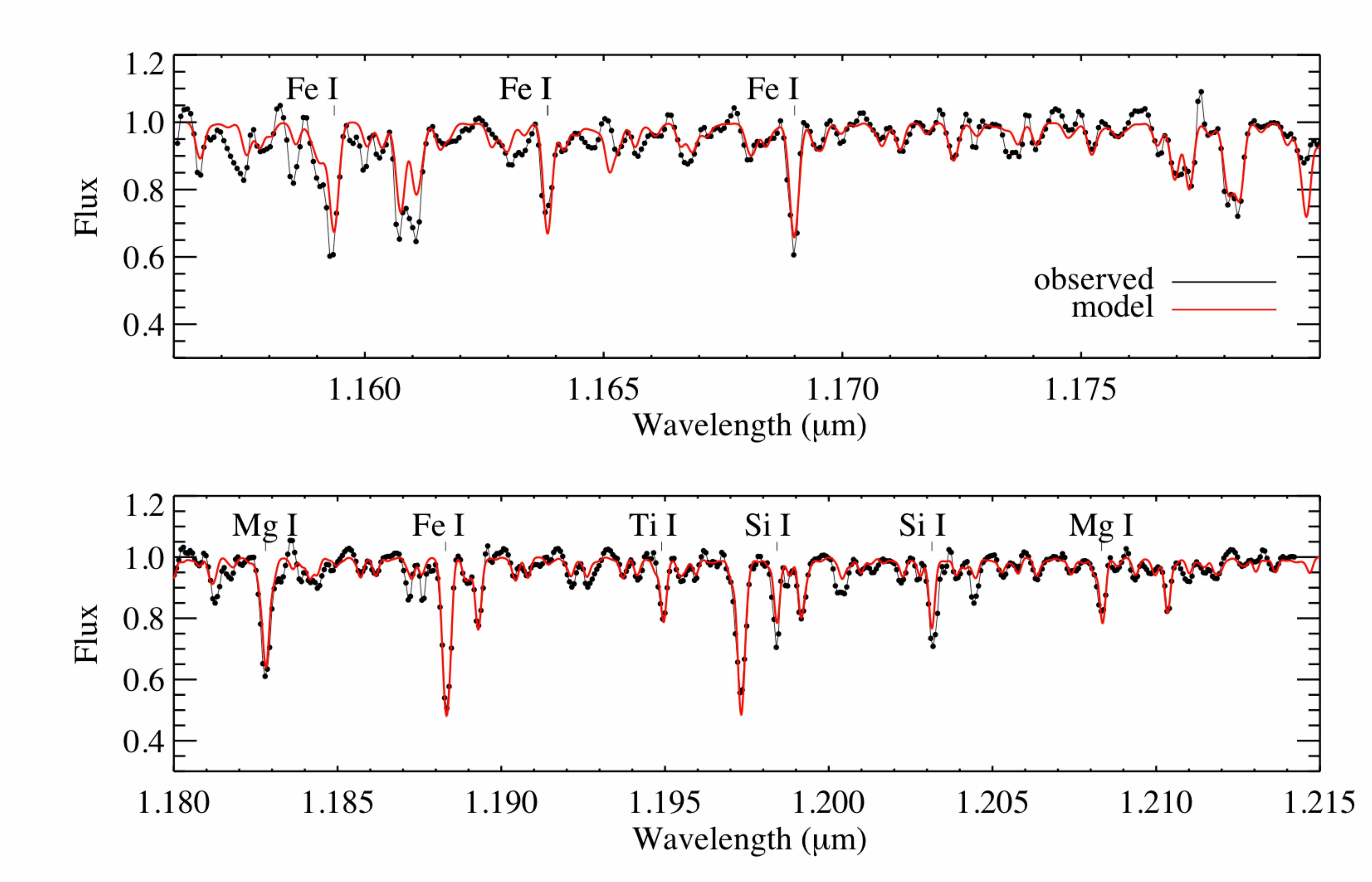}
\caption{Comparison of the observed (black) and theoretical NLTE (red) spectra
for Betelgeuse.}
\label{fig:5}   
\end{figure}

For cooler stars, such as RSG, the excitation-ionization balance method is not
useful. First, their optical spectra are too blended to allow a meaningful
estimate of a line equivalent width. Adjacent spectral lines fuse with one
another, and it is not possible to separate their individual profiles. Second,
most of the lines from singly-ionized element vanish, and there is no constraint
on the ionization balance. The method of global spectrum synthesis (Fig.
\ref{fig:5}), which relies on all spectral lines present in the observed
spectrum, is much more suitable for the analysis of RSGs. Each parameter affects
the relative strengths of the lines in subtly different ways, meaning that the
parameters may be determined via a $\chi^2$ minimisation search within a
precomputed grid of model spectra (see Gazak et al. 2014, submitted).

\begin{figure}[b]
\sidecaption
\includegraphics[scale=.4]{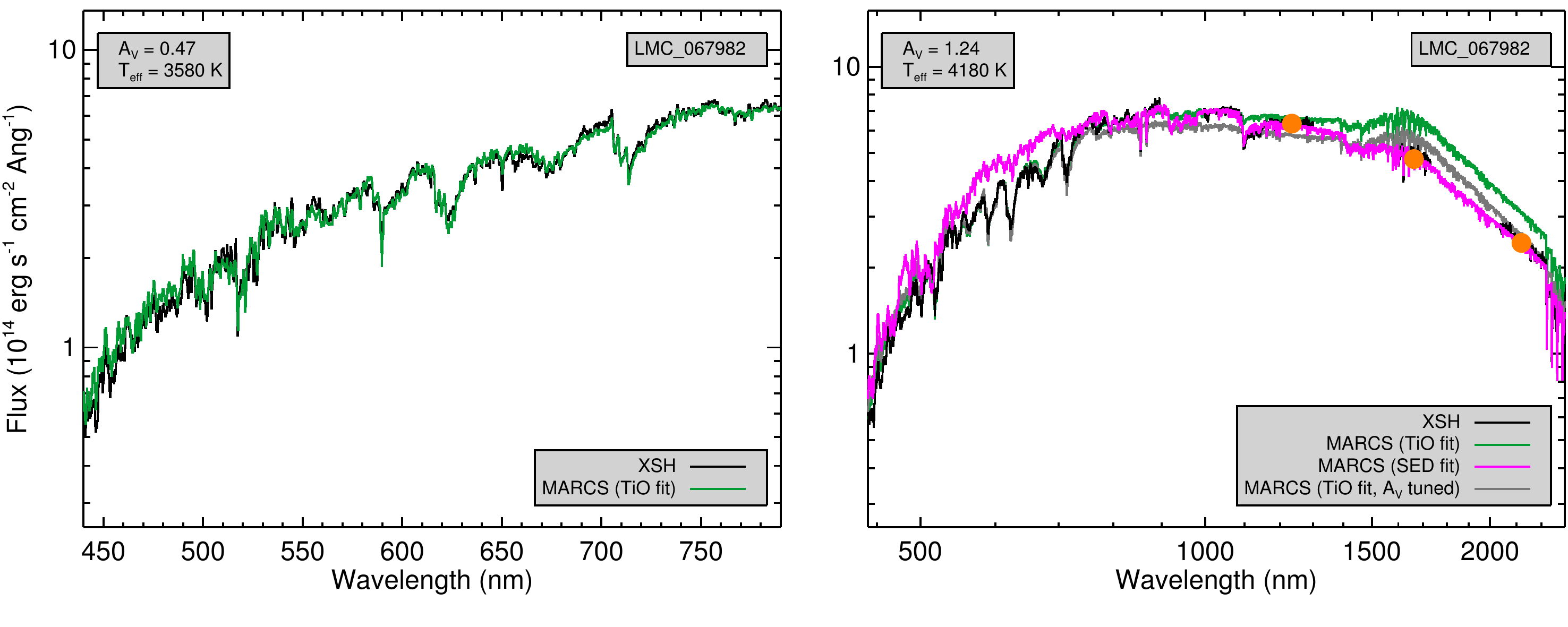}
\caption{A theoretical fit to the optical spectrum (left) and to the complete
SED for a red supergiant in LMC (Davies et al. 2013). The observed spectrum from
the X-Shooter instrument at VLT is shown with a black line. On the right plot,
the magenta and green lines correspond to the MARCS 1D LTE theoretical spectra
computed with different $\teff$. The orange symbols indicate spectro-photometric
measurements for the star.}
\label{fig:6} 
\end{figure}

\begin{figure}[b]
\sidecaption
\includegraphics[scale=.4]{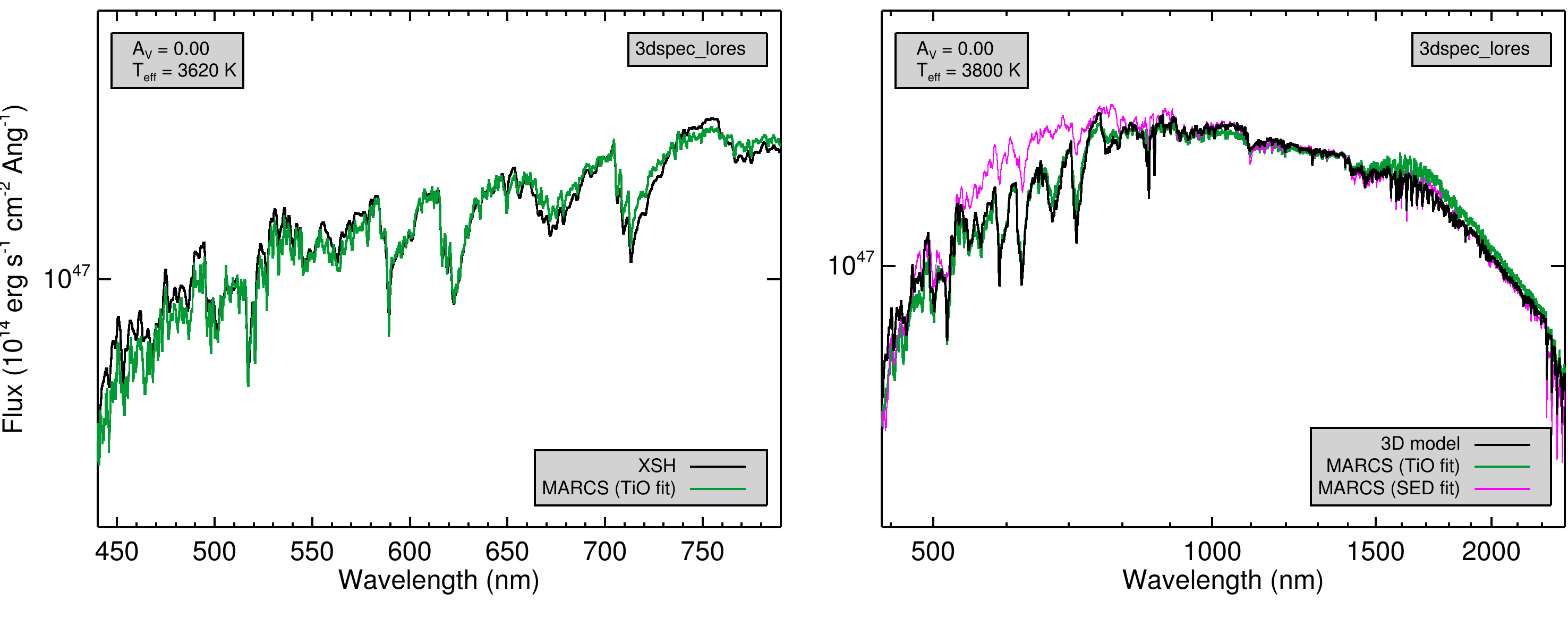}
\caption{A theoretical fit to the optical spectrum (left) and to the complete
SED (right) for a red supergiant. 3D model atmospheres (black line) improve the
consistency between the TiO bands and the IR $1.6 \mu$ diagnostics (Davies et
al. 2013).}
\label{fig:7} 
\end{figure}

The principal diagnostics of effective temperatures for RSGs have been the
strengths of the optical TiO bands which define the spectral classification
sequence. These features were modelled by \citet{2005ApJ...628..973L,
2006ApJ...645.1102L} using the MARCS model atmospheres in order to determine the
temperature scale for such stars. However, it has since been shown that, as a
consequence of deviations from hydrostatic equilibrium in RSG
atmospheres, the 1D hydrostatic models underpredict the strengths of the TiO
bands (Fig. \ref{fig:6}). This leads to effective temperatures which are too low
\citep{2013ApJ...767....3D}. Spectra computed with 3D hydrodynamical models
(Fig. \ref{fig:7}, see also Sec. \ref{sec:5}) tend to have deeper TiO absorption
bands and brighter fluxes at the minimum of opacity, $1.6 \mu$m, compared to 1D
hydrostatic models thus alleviating the problem, at least in part.

Red giants and supergiants are very useful as metallicity indicators. Their
spectra are rich in metallic lines, allowing for abundances to be determined for
a range of elements. Such work typically targets the near-IR, which coincides
with the flux peak of the stars, and which avoids the regions of the spectrum
dominated by problematic TiO bands.  More recently, this work is being adapted
so that metallicities may be retrieved at lower spectral resolutions. This is
done by isolating a region in the $J$-band (Fig. \ref{fig:5}) which is
relatively free of the molecular lines that plague the spectra of these stars,
alleviating the problem of line-blending. This means that RSGs may be studied at
distances of several Mpc, making them excellent probes of their host galaxies'
abundances \cite{2010MNRAS.407.1203D}. The atomic lines of Fe, Ti, Si and Mg
present in this region provide sufficient information to solve for the primary
free parameters in model atmospheres, i.e. effective temperature, gravity,
microturbulence and metallicity. 

\section{Physics of giant atmospheres} 
\label{sec:4}

Here we review those aspects of the physics of giants and RGSs, which can be
presently addressed with  theoretical models of atmospheres and spectral line
formation. These aspects include:
\begin{itemize}
\item{Molecular line opacities}
\item{Deviations from hydrostatic equilibrium}
\item{Deviations from local thermodynamic equilibrium}
\item{Chromospheres, MOLspheres, winds, mass loss}
\end{itemize}
For some of these aspects, such as MOLspheres\footnote{MOLsphere is a term used
for a warm optically-thick water envelope around cool stars, which has been
proposed in some studies to explain the IR observations
\citep{2001A&A...376L...1T}.} and chromospheres (see below), the models are
semi-empirical, in a sense that they are tuned to reproduce observations. 

NLTE and 3D hydrodynamical calculations can be performed based on {\it ab
initio} physical considerations. 
\subsection{Molecular line opacities}
\label{sec:5}
Until recently, one of the main limitations of spectroscopic models for giants
was the absence of accurate data for molecular line transitions. Complete TiO
linelists, which include tens of millions of transitions, have been published
only recently \citep{1998A&A...337..495P}. However, other diatomic molecules
have a significant contribution in giant spectra, including CN, CO, MgH, VO
\citep{2012A&A...544A.126D}. The major challenge is to include all these atomic
data in spectral synthesis codes and stellar atmosphere models to correctly
recover their temperature and pressure structures and predict the emergent
spectra. So far, only simplified 1D LTE hydrostatic models, like MARCS
\citep{2008A&A...486..951G} or PHOENIX \citep{1997ApJ...483..390H}, which
include accurate opacity sampling (OS) schemes, have been able to cope with the
bulky linelists. Such models are used to model RSG and RGB spectra at present.
\subsection{3D hydrodynamics and giant convective cells}
\label{sec:6}
Recently, it became possible to compute 2D and 3D radiative hydrodynamics (3D
RHD) models of stellar convection (e.g. \citealt{1998ApJ...499..914S},
\citealt{2002AN....323..213F}, \citealt{2011A&A...535A..22C}). Depending on the
properties of a star, e.g. a smaller giant or a larger supergiant, the codes may
work in two diametrally different modes. The first local setup is known as {\it
box-in-a-star} and it is usually applied to modelling solar-like stars
\citep[e.g.][]{2002AN....323..213F}. The second, global {\it star-in-a-box}
setup is applied to model very extended, variable and stochastically pulsating
stars. These are characteristics of red supergiants. The difference is
essentially that in the local setup, the equations of compressible hydrodynamics
and non-local radiation transport are solved only in a small box on a surface of
a star. In contrast, in the {\it star-in-a-box} regime, the simulation box
includes the whole star. 

The 3D RHD\footnote{See the review lecture on 3D NLTE spectroscopy in the book.}
models are much more successful than classical 1D hydrostatic models in
reproducing a wealth of observational information, including the line shapes and
bisectors, center-to-limb variation, and brightness contrast 
\citep{2005ARA&A..43..481A}. However, such models are exceptionally
sophisticated and can not be computed for a large range of stellar parameters
due to prohibitively long computation times. At present, the use of 3D RHD
models is restricted to the analysis of individual stars, such as the bright
metal-poor giant HD 122563 
\citep{collet09} and the red supergiant Betelgeuse \citep{2010A&A...515A..12C}.
The 3D RHD models are also used to explore the validity range of simpler 1D
hydrostatic model atmospheres 
\citep[e.g.][]{2007A&A...469..687C}.

\begin{figure}[b]
\sidecaption
\includegraphics[scale=.3]{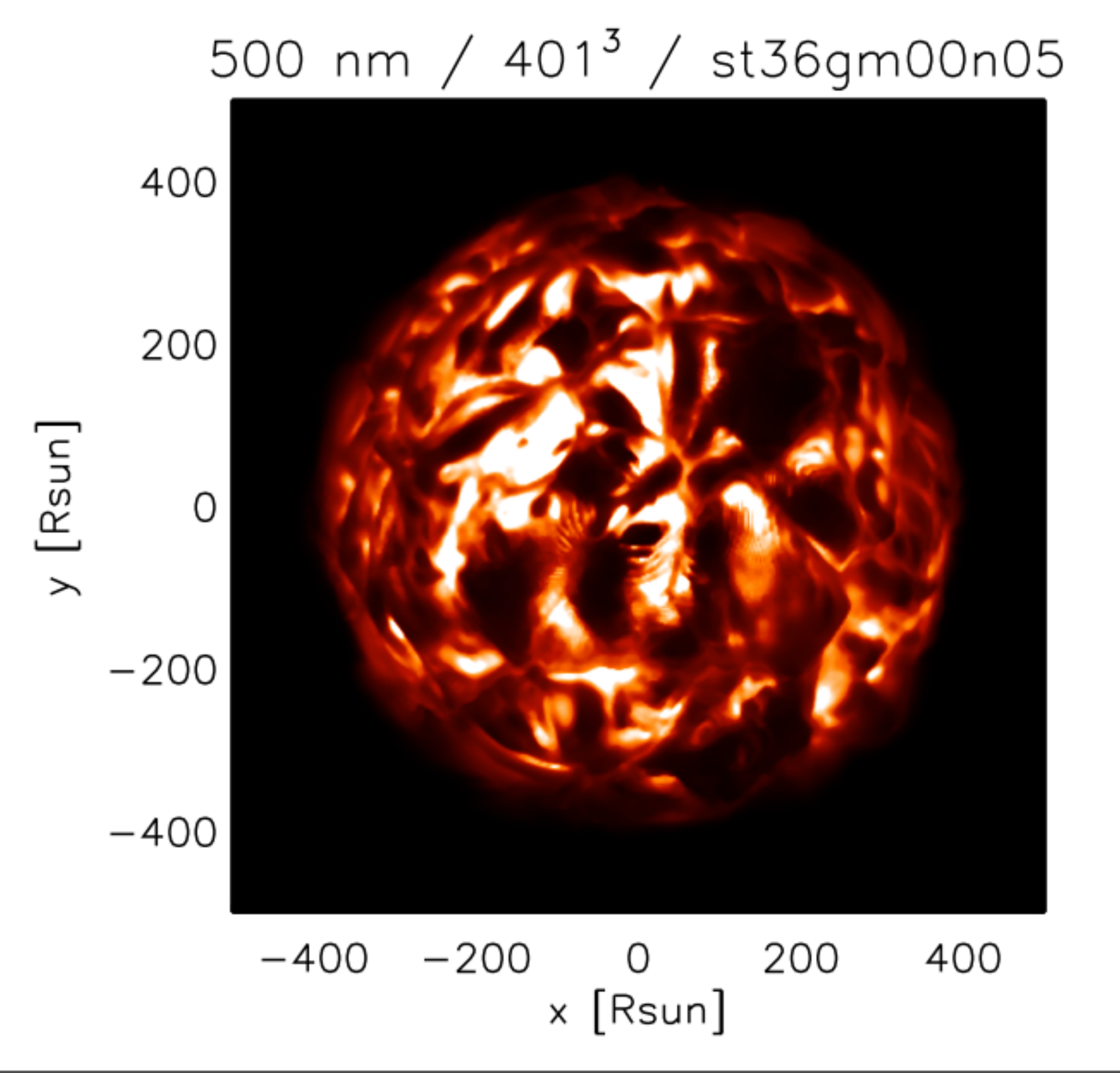}
\caption{Emergent intensity at $500$ nm from a 3D RHD simulation of convection
for a red supergiant Betelgeuse. The model is characterised by $\teff = 3710$ K,
$\logg = 0.047$, and the solar metallicity. The cubic grid contains 401$^3$
nodes \cite[][]{2011A&A...535A..22C}.}
\label{fig:8}   
\end{figure}

Fig. \ref{fig:8} shows a snapshot from a star-in-a-box simulation of convection
on Betelgeuse, as computed with the CO5BOLD code \citep{2011A&A...535A..22C}.
The emergent intensity is shown at the wavelength of $500$ nm. Huge giant
convection cells predicted by the model are consistent with imaging observations
of the star from space, e.g from the Hubble Space Telescope
\citep{1996ApJ...463L..29G}.

\subsection{NLTE}
\label{sec:6}
Another major complication in the analysis of spectra of cool giants is to
include departures from Local Thermodynamic Equilibrium (LTE). As a consequence
of very low surface densities, $-0.5 < \log g < 3$, collisions between particles
are too weak to establish local equilibrium in a stellar atmosphere. Each parcel
of the gas is influenced by photons originating elsewhere. This is what we call
- \textit{non-local radiation field}, and thus non-local thermodynamic
equilibrium. Research in NLTE physics of cool giant  and supergiant atmospheres
is still at its infancy, however, it is clear already now that spectral lines of
different atoms and molecules do not form under LTE conditions.

\begin{figure}[b]
\sidecaption
\includegraphics[scale=.35]{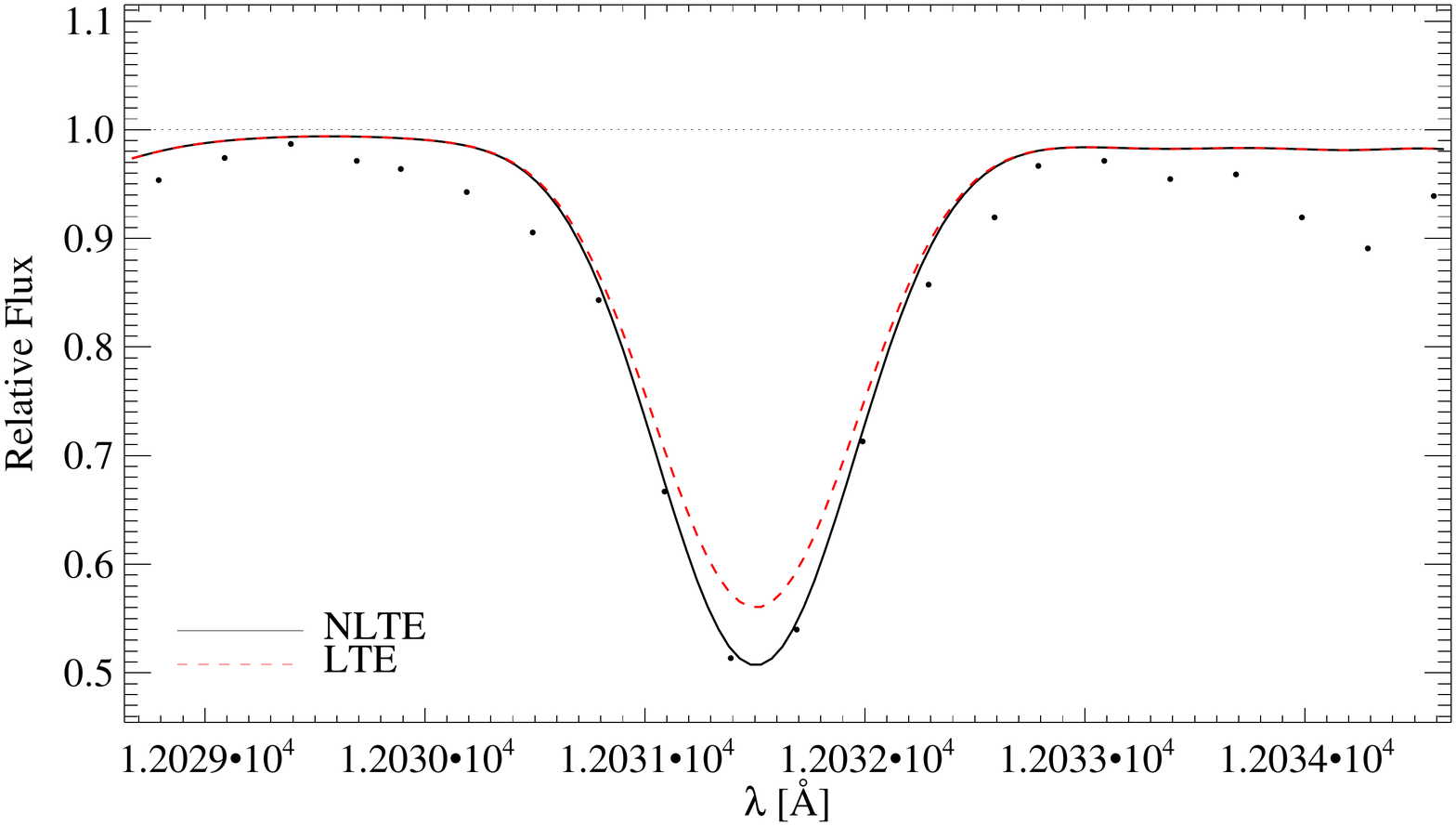}
\caption{Subaru$/$IRCS high resolution observations (black dots) J-band
observations of Si I lines in the spectrum of the Per OB1 RSG HD 14270 (12031
$\AA$) compared with a LTE (red, dashed) and a NLTE (black, solid) fit. The fit
profiles have been calculated with $\teff = 3800$ K, $\log g=1.0$,$[Z]=0.0$, and
microturbulence $Vmic=5$km/s \citep{2013ApJ...764..115B}.}
\label{fig:9} 
\end{figure}

\cite{2012MNRAS.427...27B, 2012ApJ...751..156B, 2013ApJ...764..115B} proved from
{\it ab initio} statistical equilibrium calculations, that ions such as Fe I, Ti
I, Si I, Cr I, Ti I, are very sensitive to strong photo-ionising radiation
fields. In this case, the effect of NLTE is to reduce the number densities of
atoms in a minority ionisation stage, thus for a given abundance the line
equivalent width is smaller than that given by LTE. This effect is larger in
metal-poor stars. At solar metallicity, the opposite behaviour is encountered
sometimes. For example, the IR transitions of Si I and Ti I show {\it negative}
NLTE effects such that the LTE abundance is higher than the NLTE one (Fig.
\ref{fig:9}). According to recent studies, the LTE and NLTE abundances may
differ by an order of magnitude, from $-0.5$ to $+1.0$ dex, depending on the
element and ionisation stage, intrinsic line parameters (wavelength, level
excitation potential), and stellar parameters. 

The NLTE analysis of molecular spectral lines is an unexplored territory. New
studies show the effect of NLTE in is to cause stronger molecular absorption
bands that leads to extra cooling in the outer atmospheric layers
\citep{2013EAS....60..111L}.

Detailed theoretical and observational studies are necessary to understand how
NLTE influences the estimates of stellar parameters and abundances from giant
spectra.
\subsection{Chromospheres, MOLspheres, winds, and mass loss}
\label{sec:7}

\begin{figure}[b]
\sidecaption
\includegraphics[scale=.50]{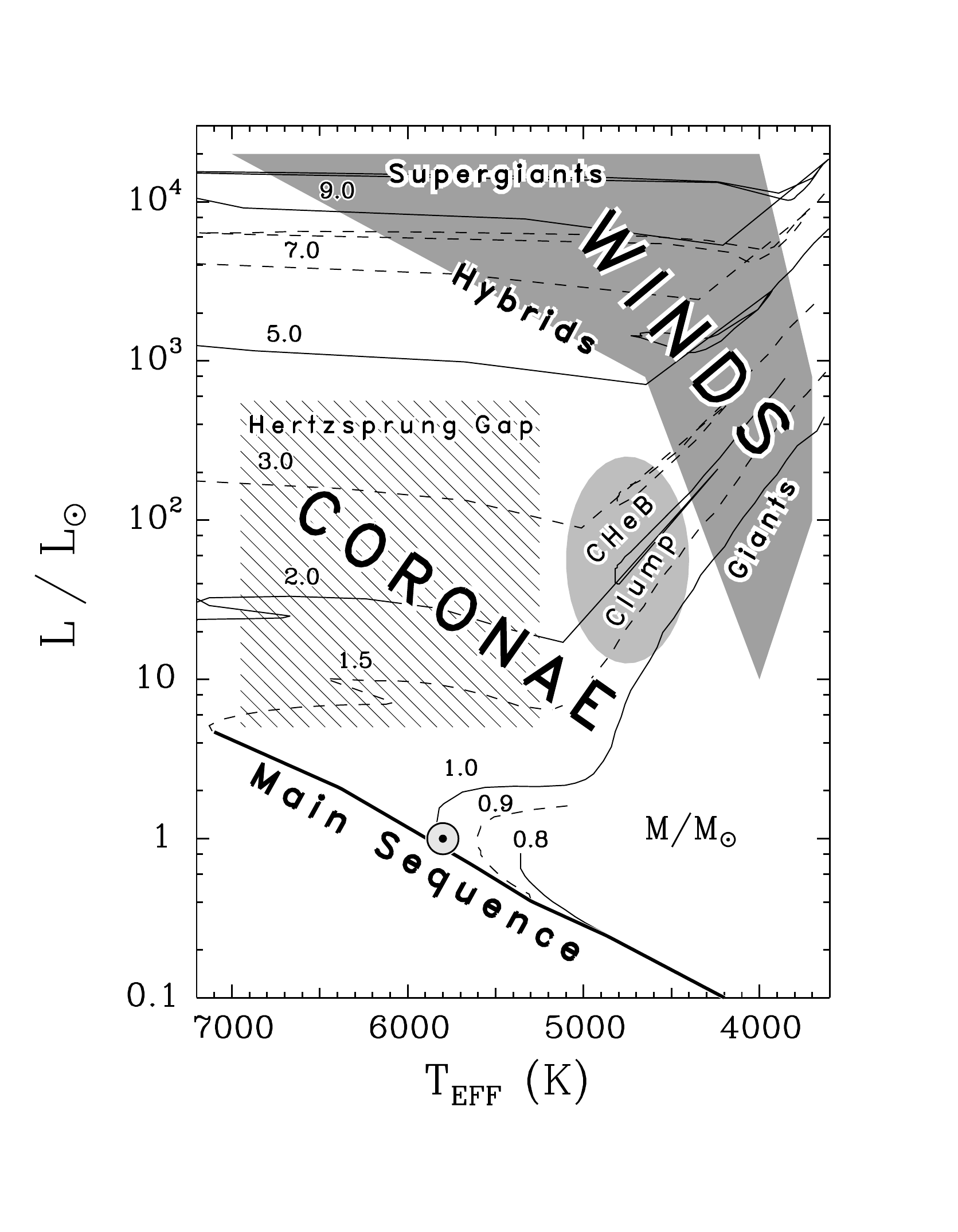}
\caption{An H-R diagram of late-type stars \citep[][]{2010MmSAI..81..553A}.
Stellar winds and chromospheres are typical for cool giants and supergiants.}
\label{fig:10} 
\end{figure}

Fig. \ref{fig:10} shows the late-type stars in the H-R diagram
\citep{2010MmSAI..81..553A}. The illustration is similar to the
\cite{1979ApJ...229L..27L} diagram of \textit{solar} and \textit{non-solar} type
stars; the former with chromospheres, transition regions, and coronae, and the
latter with chromospheres only. It is thought that the absence of a corona is a
consequence of strong winds \citep[see also][]{2011A&A...533A.107D}.

Chromospheres are easy to detect through observations. For instance,
high-quality spectroscopic observations of RGB tip stars in globular clusters
produced compelling evidence for chromospheric activity, based on the emission
wings of $H_{\alpha}$ line at $656.3$ nm \citep{2008AJ....135.1117M}. Accoridng
to models, the wings form in the chromosphere\footnote{Interestingly, only stars
with $log(L/L_\odot) \geq 2.4$ show emission wings.}, although they are also
affected by stellar pulsations. Other diagnostic features sensitive to activity
are the UV Ca H and K lines at $395$ nm and the Mg II resonance lines at at
$280$ nm \citep[e.g.][]{2013ApJ...772...90L}. These strong lines develop
chromospheric emission in the innermost cores. Since stellar rotation, and thus
chromospheric activity, declines with the age of a star, the strength of the
line core emission is a useful measure of stellar ages
\citep{1991ApJ...375..722S}. Strong correlation was also discovered between the
strengths of Mg II resonance lines and TiO molecular bands
\citep{1985ApJ...291L..51S}. 

The inability to fit some features in the IR spectra of cool giants with 1D
hydrostatic LTE models gave rise to   extensions of the classical models. Such
extended models, for example, include MOLsphere, a very cool molecular layer
above the photosphere \citep{2001A&A...376L...1T, 2008A&A...489.1271T}. The
models assume  a very low temperature of $1000 - 2000$ K and the column density
of molecules is adjusted to fit the observed spectra. It is unclear whether the
appeal to such ad-hoc extensions is justified given the major uncertainties in
the analysis of molecular spectral lines caused by the limitations of 1D
hydrostatic models with LTE.

Red giants and supergiants lose mass through low-speed winds of few tens of
km$/$s. The rate of mass loss can be very high, for AGB stars up to $10^{-5}
M_{\odot}/$year and for RSGs up to $10^{-4} M_{\odot}/$year
\citep{2010A&A...523A..18D}. In- and out-flow motions are 'seen' in the
high-resolution spectra of giants: the most pronounced features are shifts and
asymmetries of strong line cores, like $H_{\alpha}$. Mass outflows are
responsible for the circumstellar shells that cause IR emission in the stellar
spectra. Some insight into the mass loss phenomenon can also be gained from
interferometry. It has been suggested that RSG winds contribute to larger radii
at $712$ nm as measured by interferometric techniques. 

Models of oxygen-rich stars with effective temperatures below $3000$ K must
include dust formation \citep{2001ApJ...557..798F}. The effect of the latter is
two-fold: dust condensation warms up the atmosphere and decreases the abundance
of elements locked up in the gas phase. Different elements, such as Ti, Zr, and
Al, condense into dust grains. For an observer, the most relevant is the change
in the number density of TiO, which can be detected in the spectra. The impact
of more sophisticated models on the abundance analysis and stellar parameter
determinations remains to be quantified.
\section{Conclusions}\label{sec:8}
To conclude, modelling spectra of cool giants and supergiants is, doubtlessly, a
very challenging problem. Different physical phenomena must be included in the
atmosphere and line formation models to allow for an accurate determination of
surface parameters: molecular line opacities, deviations from 1D hydrostatic
equilibrium and from local thermodynamic equilibrium, chromospheres, winds, and
mass loss. We have attained the necessary minimum level of complexity to address
the simpler problems, like the determination of metallicity and bolometric flux.
However, more detailed investigations, like the analysis of TiO or CO molecular
bands and the interpretation of strong lines like H$_\alpha$, Ca H $\&$ K, Mg
II, which contain the essential physics, shall await for more complex models
including the above-mentioned effects. In the view of the enormous perspectives
offered by observing giants and RSGs throughout the Milky Way and in other
galaxies, the need for more physically-realistic models is well-justified.
\begin{acknowledgement}
Figures from the following papers have been reproduced with permission
from the authors and from the publishers (c) ESO: Chiavassa et al. A
\&A, 535, A22, 2011; Decin et al. A\&A, 548, A113, 2012; Georgy et
al. A\&A, 558, A103, 2013; Lancon \& Wood, A\&AS, 146, 217, 2000;
Kervella et al., A\&A, 504, 115, 2009. Figures from the following
papers have been reproduced by permission of the AAS: Davies et al.,
ApJ, 767:3, 2013. This work was partly supported by the European Union FP7
programme through ERC grant number 320360.

\end{acknowledgement}
\bibliographystyle{spbasic}
\bibliography{references}
\end{document}